\documentclass[12pt]{iopart}

\usepackage{graphicx}

\begin{document}

\title[Renormalization group structure for sums generated by incipiently chaotic maps]{Renormalization group structure for sums of variables generated by incipiently chaotic maps}

\author{Miguel Angel Fuentes$^{1,2,3}$ and Alberto Robledo$^{4}$}

\address{$^1$Santa Fe Institute, 1399 Hyde Park Road, Santa Fe, New Mexico 87501, USA}
\address{$^2$Centro At\'omico Bariloche, Instituto Balseiro and CONICET, 8400 Bariloche, Argentina}
\address{$^3$ Center for Advanced Studies in Ecology and Biodiversity, Facultad de Ciencias Biol\'ogicas, Pontificia Universidad Cat\'olica de Chile, Casilla 114-D, Santiago CP 6513677, Chile}
\address{$^4$Instituto de F\'{\i}sica, Universidad Nacional Aut\'onoma de M\'exico,
Apartado Postal 20-364, M\'exico 01000 DF, Mexico}

\ead{fuentesm@santafe.edu}
\begin{abstract}
We look at the limit distributions of sums of deterministic chaotic
variables in unimodal maps and find a remarkable renormalization group (RG)
structure associated to the operation of increment of summands and
rescaling. In this structure - where the only relevant variable is the
difference in control parameter from its value at the transition to chaos -
the trivial fixed point is the Gaussian distribution and a novel nontrivial
fixed point is a multifractal distribution that emulates the Feigenbaum
attractor, and is universal in the sense of the latter. The crossover
between the two fixed points is explained and the flow toward the trivial
fixed point is seen to be comparable to the chaotic band merging sequence.
We discuss the nature of the Central Limit Theorem for deterministic
variables.

\end{abstract}

\maketitle

\section{Introduction}

As it is well documented \cite{kaminska1}, increasingly larger sums of
iterates of chaotic mappings give rise to a Gaussian stationary distribution
in the same way independent random variables do according to the ordinary
central limit theorem \cite{vankampen1, khinchin1}. This deep-seated
property, remarkably shared by deterministic and random systems composed of
essentially uncorrelated variables, naturally raises questions about the
existence, and if so, uniqueness or diversity, of limit distributions for
systems made up of deterministic correlated variables. Related to these
issues recent \cite{tsallis1}-\cite{grassberger1} numerical explorations of
time averages of iterates at the period-doubling transition to chaos \cite{schuster1, Hu} have been presented. Since the trajectories linked to this
critical attractor are nonergodic and nonmixing the question of whether
there is such stationary distribution for sums of iterates at the transition
to chaos holds added interest as it may provide new angles to appraise the
statistical mechanical analogy that is found in chaotic dynamics \cite%
{robledo1}.

The dynamics toward and at the Feigenbaum attractor is now known in much
detail \cite{robledo1, robledo2}, therefore, it appears feasible to analyze
also the properties of sums of iterate positions for this classic nonlinear
system with the same kind of analytic reasoning and numerical thoroughness.
Here we present the results for sums of chronological positions of
trajectories associated to\ quadratic unimodal maps. We consider the case of
the sum of positions of trajectories inside the Feigenbaum attractor as well
as those within the chaotic $2^{K}$-band attractors obtained when the
control parameter is shifted to values larger than that at the transition to
chaos. Time and ensemble averages differ at the transition to chaos and here
we chose to study the time average of a single trajectory initiated inside
the attractor since all such trajectories, as explained below, are simply
related. Clearly, time and ensemble averages are equivalent for chaotic
attractors. From the information obtained we draw conclusions on the
properties of the stationary distributions for these sums of variables. Our
results, that reveal a multifractal stationary distribution that mirrors the
features of the Feigenbaum attractor, can be easily extended to other
critical attractor universality classes and other routes to chaos. About the relevance of our findings to physical systems, it is interesting to note, as one example, the parallels that have been found to exist between the dynamics at the noise-perturbed onset of chaos in unimodal maps and the
dynamics of glass formation \cite{Brobledo}. In this connection chaotic band
merging plays a central role in the relaxation properties of time correlation
functions, while the multifractal attractor and multiband attractors in its
neighborhood display the characteristic aging scaling property of glass formers \cite{Brobledo}.

The overall picture we obtain is effectively described within the framework
of the renormalization group (RG) approach for systems with scale invariant
states or attractors. Firstly, the RG transformation for the distribution of
a sum of variables is naturally given by the change due to the increment of
summands followed by a suitable restoring operation. Second, the limit
distributions can be identified as fixed points reached according to whether
the acting relevant variables are set to zero or not. Lastly, the
universality class of the non-trivial fixed-point distribution can be
assessed in terms of the existing set of irrelevant variables.

As it is well known \cite{schuster1, Hu} a few decades ago the RG approach was
successfully applied to the period-doubling route to chaos displayed by
unimodal maps. In that case the RG transformation is \textit{functional
composition} and rescaling of the mapping and its effect re-enacts the
growth of the period doubling cascade. In our case the RG transformation is
the \textit{increment of terms} and adjustment of the sum of positions and
its effect is instead to go over again the merging of bands in the chaotic
region.

Specifically, we consider the Feigenbaum map $g(x)$, obtained from the fixed
point equation $g(x)=\alpha g(g(x/\alpha ))$ with $g(0)=1$ and $g^{\prime
}(0)=0$, and where $\alpha =-2.50290...$ is one of Feigenbaum's universal
constants \cite{schuster1, Hu}. (For expediency we shall from now on denote the
absolute value $\left\vert \alpha \right\vert $ by $\alpha $). Numerically,
the properties of $g(x)$ can be conveniently obtained from the logistic map $%
f_{\mu }(x)=1-\mu x^{2},\;-1\leq x\leq 1$, with $\mu =$ $\mu _{\infty
}=1.401155189092..$. The dynamics associated to the Feigenbaum map is
determined by its multifractal attractor. For a recent detailed description
of these properties see \cite{robledo1, robledo2}. For values of $\mu >$ $%
\mu _{\infty }$ we employ a well-known scaling relation supported by
numerical results.

Initially we present properties of the sum of the absolute values $%
\left\vert x_{t}\right\vert $ of positions $x_{t}=f_{\mu _{\infty
}}(x_{t-1}),\;t=1,2,3,...$, as a function of total time $N$ visited by the
trajectory with initial position $x_{0}=0$, and obtain a patterned linear
growth with $N$. We analyze this intricate fluctuating pattern, confined
within a band of finite width, by eliminating the overall linear increment
and find that the resulting stationary arrangement exhibits features
inherited from the multifractal structure of the attractor. We derive an
analytical expression for the sum that corroborates the numerical results
and provide an understanding of its properties. Next, we consider the
straight sum of $x_{t}$, where the signs taken by positions lessen the
growth of its value as $N$ increases and the results are consistently
similar to those for the sum of $\left\vert x_{t}\right\vert $, i.e. linear
growth of a fixed-width band within which the sum displays a fluctuating
arrangement. Numerical and analytical details for the sum of $x_{t}$ are
presented. Then, we show numerical results for the sum of iterated positions
obtained when the control parameter is shifted into the region of chaotic
bands. In all of these cases the distributions evolve after a characteristic
crossover towards a Gaussian form. Finally, we rationalize our findings in
terms of an RG framework in which the action of the Central Limit Theorem
plays a fundamental role and provide details of the crossover from multiband
distributions to the gaussian distribution. We discuss our results.

\section{Sums of positions at the chaos threshold}

\subsection{Sum of absolute values $\left\vert x_{t}\right\vert $}

The starting point of our study is the evaluation of
\begin{equation}
y_{\mu }(N)\equiv \sum\limits_{t=1}^{N}\left\vert x_{t}\right\vert ,
\label{sumabs1}
\end{equation}%
with $\mu =\mu _{\infty }$ and with $x_{0}=0$. Fig. 1A shows the result,
where it can be observed that the values recorded, besides a repeating
fluctuating pattern within a narrow band, increase linearly on the whole.
The measured slope of the linear growth is $c=0.56245...$ Fig. 1B shows an
enlargement of the band, where some detail of the complex pattern of values
of $y_{\mu _{\infty }}(N)$ is observed. A stationary view of the mentioned
pattern is shown in Fig. 1C, where we plot%
\begin{equation}
y_{\mu _{\infty }}^{\prime }(N)\equiv \sum\limits_{t=1}^{N}\left( \left\vert
x_{t}\right\vert -c\right) ,  \label{sumabs2}
\end{equation}%
in logarithmic scales. There, we observe that the values of $y_{\mu _{\infty
}}^{\prime }(N)$ fall within horizontal bands interspersed by gaps,
revealing a fractal or multifractal set layout. The top (zeroth) band
contains $y_{\mu _{\infty }}^{\prime }$ for all the odd values of $N$, the
1st band next to the top band contains $y_{\mu _{\infty }}^{\prime }$ for
the even values of $N$ of the form $N=2+4m$, $m=0,1,2,...$ The 2nd band next
to the top band contains $y_{\mu _{\infty }}^{\prime }(N)$ for $%
N=2^{2}+2^{3}m$, $m=0,1,2,...$, and so on. In general, the $k$-th band next
to the top band contains $y_{\mu _{\infty }}^{\prime }(2^{k}+2^{k+1}m)$, $%
m=0,1,2,...$ Another important feature in this figure is that the $y_{\mu
_{\infty }}^{\prime }(N)$ for subsequences of $N$ each of the form $%
N=(2l+1)2^{k}$, $k=0,1,2,...$, with $l$ fixed at a given value of $%
l=0,1,2,...$, appear aligned with a uniform slope $s=-1.323...$ The parallel
lines formed by these subsequences imply the power law $y_{\mu _{\infty
}}^{\prime }(N)\sim N^{s}$ for $N$ belonging to such a subsequence.

\begin{figure}[h]
\centering \includegraphics[width=12cm,angle=0]{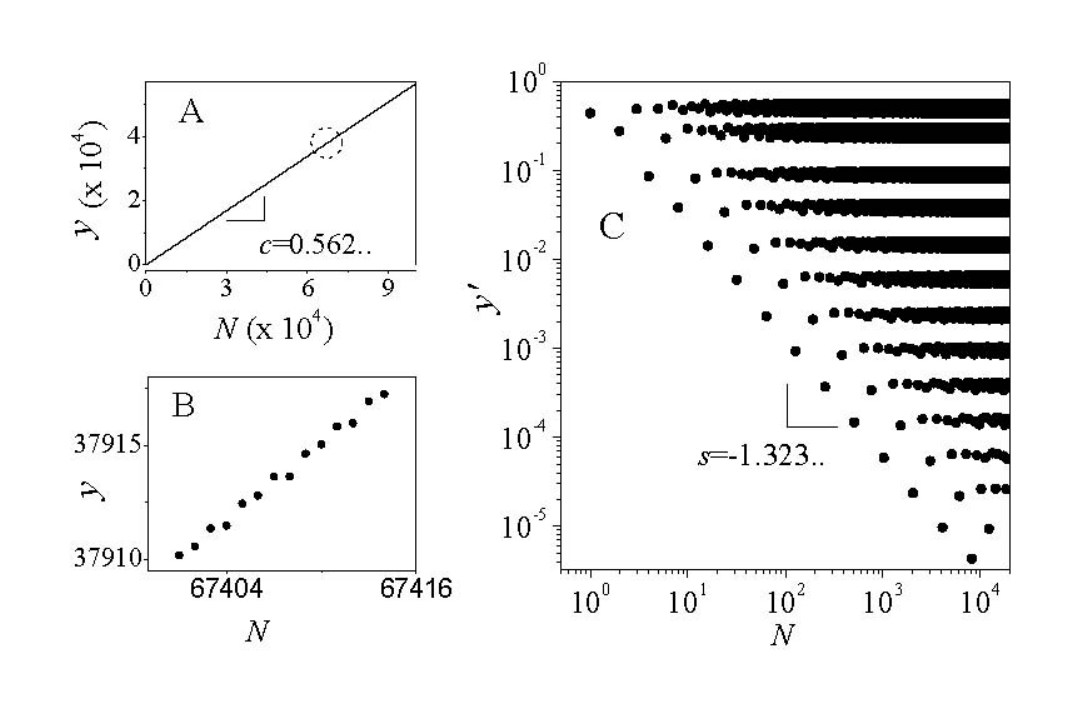}
\caption{A) Sum $y_{\mu _{\infty }}(N)$ of absolute values of visited points $%
x_{t}$, $t=0,...,N$, of the Feigenbaum's attractor with initial condition $%
x_{0}=0$. B) A closer look of the path of the sum (see dotted circle in A),
for values of $N$ around 67410. C) Centered sum $y_{\mu _{\infty }}^{\prime
}(N)$ in logarithmic scales. See text.}
\label{f1}
\end{figure}

\begin{figure}[tbp]
\centering \includegraphics[width=10cm,angle=0]{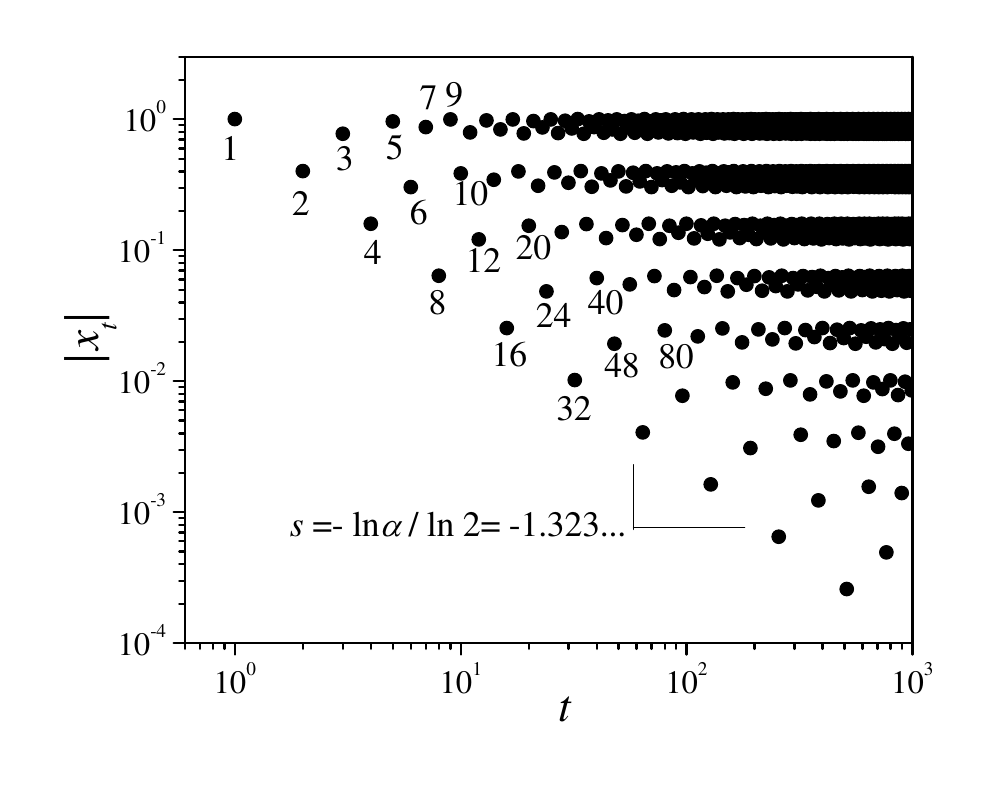}
\caption{Absolute value of trajectory positions $x_{t}$, $t=0,...$, for the
logistic map $f_{\mu }(x)$ at $\mu _{\infty }$, with initial condition $%
x_{0}=0$, in logarithmic scale as a function of the logarithm of the time $t$%
, also shown by the numbers close to the points.}
\label{f2}
\end{figure}

It is known \cite{robledo1, robledo3} that these two characteristics of $%
y_{\mu _{\infty }}^{\prime }(N)$ are also present in the layout of the
absolute value of the individual positions $\left\vert x_{t}\right\vert $, $%
t=1,2,3,...$ of the trajectory initiated at $x_{0}=0$; and this layout
corresponds to the multifractal geometric configuration of the points of the
Feigenbaum's attractor, see Fig. 2. In this case, the horizontal bands of
positions separated by equally-sized gaps are related to the set of
period-doubling `diameters' \cite{schuster1, Hu} employed for the construction
of the multifractal \cite{robledo2}. The identical slope shown in the
logarithmic scales by all the position subsequences $\left\vert
x_{t}\right\vert $, $t=(2l+1)2^{k}$, $k=0,1,2,...$, each formed by a fixed
value of $l=0,1,2,...$, implies the power law $\left\vert x_{t}\right\vert
\sim t^{s}$, $s=-\ln \alpha /\ln 2=-1.3236...$, as the $\left\vert
x_{t}\right\vert $ can be expressed as $\left\vert x_{t}\right\vert \simeq
\left\vert x_{2l+1}\right\vert \ \alpha ^{-k},t=(2l+1)2^{k}$, $k=0,1,2,...$,
or, equivalently, $\left\vert x_{t}\right\vert \sim t^{s}$. Notice that the
index $k$ also labels the order of the bands from top to bottom. The power
law behavior involving the universal constant $\alpha $ of the subsequence
positions reflect the approach of points in the attractor toward its most
sparse region at $x=0$ from its most compact region, as the positions at odd
times $\left\vert x_{2l+1}\right\vert =x_{2l+1}$, those in the top band,
correspond to the densest region of the set.

Having uncovered the through manifestation of the multifractal structure of
the attractor into the sum $y_{\mu _{\infty }}^{\prime }(N)$ we proceed to
derive this property and corroborate the numerical evidence. Consider Eq. (%
\ref{sumabs1}) with $N=2^{k}$, $k=0,1,2,...$, the special case $l=0$\ in the
discussion above. Then the numbers of terms $\left\vert x_{t}\right\vert $
per band in $y_{\mu _{\infty }}(2^{k})$ are: $2^{k-1}$ in the top band\ ($%
j=0 $), $2^{k-2}$ in the next band\ ($j=1$),..., $2^{0}$ in the ($k-1$)-th
band, plus an additional position in the $k$-th band. If we introduce the
average of the positions on the top band%
\begin{equation}
\left\langle a\right\rangle \equiv 2^{-(k-1)}\sum_{j=0}^{2^{k-1}}x_{2j+1},
\label{ave1}
\end{equation}%
the sum $y_{\mu _{\infty }}(2^{k})$ can be written as%
\begin{equation}
y_{\mu _{\infty }}(2^{k})=\left\langle a\right\rangle
2^{k-1}\sum_{j=0}^{k-2}(2\alpha )^{-j}+\alpha ^{-(k-1)}+\alpha ^{-k}.
\label{sumabs3}
\end{equation}%
Doing the geometric sum \ above and expressing the result as $y_{\mu
_{\infty }}(2^{k})=c2^{k}+d\alpha ^{-k}$, we have%
\begin{equation}
c=\frac{\left\langle a\right\rangle \ \alpha }{2\alpha -1},~~~
d=\left( 1-\frac{\left\langle a\right\rangle \ 2\alpha }{2\alpha -1}\right)
\ \alpha +1\ .  \label{slopeandshift1}
\end{equation}%
Evaluation of Eq. (\ref{ave1}) yields to $\left\langle a\right\rangle
=0.8999...$, and from this we obtain $c=0.56227...$ and $d=0.68826...$ We
therefore find that the value of the slope $c$ in Fig. 1A is properly
reproduced by our calculation. Also, since $\ln \left[ y_{\mu _{\infty
}}(2^{k})-c2^{k}\right] =\ln d-k\ln \alpha $, or, equivalently, $\ln y_{\mu
_{\infty }}^{\prime }(N)=\ln d-N\ln \alpha /\ln 2$,$\;N=2^{k}$,$%
\;k=0,1,2,... $, we corroborate that the value of the slope $s$ in the inset
of Fig. 1C is indeed given by $s=-\ln \alpha /\ln 2=1.3236...$ (We have made
use of the identity $\alpha ^{-k}=N^{-\ln \alpha /\ln 2}$,$\;N=2^{k}$,$%
\;k=0,1,2,...$).

\subsection{Sum of values of $x_{t}$}

\begin{figure}[tbp]
\centering \includegraphics[width=11cm,angle=0]{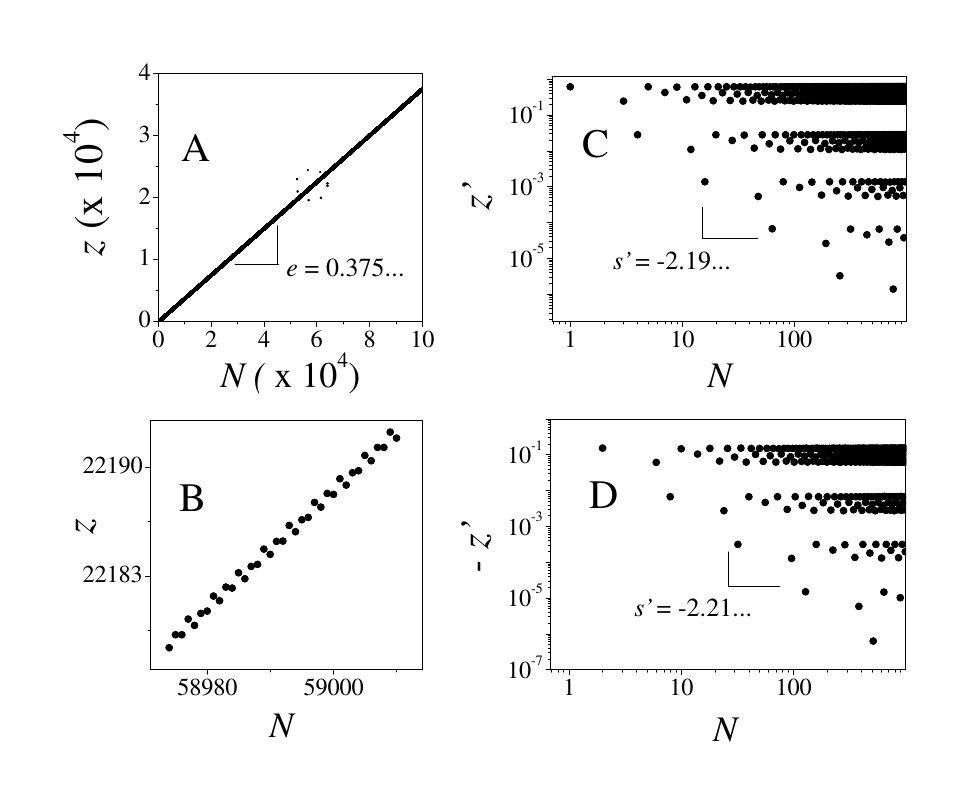}\newline
\caption{A) Sum $z_{\mu _{\infty }}(N)$ of values of visited points $x_{t}$, $%
t=0,...,N$, of the Feigenbaum's attractor with initial condition $x_{0}=0$.
B) A closer look of the path of the sum (see dotted circle in A), for values
of $N$ around 59000. C and D) Centered sum $z_{\mu _{\infty }}^{\prime }(N)$
in logarithmic scales. See text.}
\label{f3}
\end{figure}

When considering the signs taken by positions $x_{t}$ we note that their sum,
\begin{equation}
z_{\mu }(N) \equiv \sum _{t=0}^{N}x_{t},  \label{sumnorm1}
\end{equation}
when $N=2^{k}$, $\mu =$ $\mu _{\infty }$ and $x_{0}=0$, can be immediately
obtained from the above derivation for $y_{\mu _{\infty }}(2^{k})$ simply by
replacing $\alpha ^{-j}$ by $(-1)^{j}\alpha ^{-j}$, as the $x_{t}$ of
different signs of the trajectory starting at $x_{0}=0$ fall into separate
alternating bands (described above and shown in Fig. 2). In short, $%
x_{t}\simeq (-1)^{j}x_{2l+1}\ \alpha ^{-j}$, $t=(2l+1)2^{k}$, $k=0,1,2,...$
Writing $z_{\mu _{\infty }}(2^{k})$ as $z_{\mu _{\infty
}}(2^{k})=e2^{k}+f(-1)^{k}\alpha ^{-k}$, we have%
\begin{equation}
e=\frac{\left\langle a\right\rangle \alpha }{2\alpha +1},~~~
f=\left( 1-\frac{\left\langle a\right\rangle 2\alpha }{2\alpha +1}\right)
(-\alpha )+1.  \label{slopeandshift2}
\end{equation}%
Use of $\left\langle a\right\rangle =0.8999...$, leads to $e=0.37503...$ and
$f=0.37443...$ Since $\ln \left( z_{\mu _{\infty }}(2^{k})-e2^{k}\right)
=\ln f-k\ln \alpha $, or $\ln z_{\mu _{\infty }}^{\prime }(N)=\ln f-N\ln
\alpha /\ln 2,\;N=2^{k},\;k=0,1,2,...$, where $z_{\mu _{\infty }}^{\prime
}(2^{k})\equiv $\ $z_{\mu _{\infty }}(2^{k})-e2^{k}$, we obtain for the
value of the slope $s^{\prime }$ associated to the plot of $z_{\mu _{\infty
}}^{\prime }(N)\sim N^{s^{\prime }}$ in logarithmic scales the number $%
s^{\prime }=-\ln (\alpha ^{2}-1)\ln \alpha /\ln 2=-2.1984...$(where the
factor $\ln (\alpha ^{2}-1)$ takes into account the fact that consecutive
values of the same sign in $z_{\mu _{\infty }}^{\prime }(2^{k})$ have $k=2m$
or $2m+1$, $m=0,1,2,...$) Our numerical evaluations for $z_{\mu _{\infty
}}(N)$ and $z_{\mu _{\infty }}^{\prime }(N)$, shown in Fig. 3, reproduce the
values given above for the slopes $e$ and $s^{\prime }$. Therefore, our
numerical and analytical results are in agreement also in this case.

The relationship between the sum $y_{\mu _{\infty }}(N)$ with initial
condition $x_{0}=0$ and all other sums $Y_{\mu _{\infty }}(M)$ of
consecutive positions with any initial condition $x_{0}$ inside the
attractor can be obtained by inspection of Fig. 2. The sum $Y_{\mu _{\infty
}}(M)$ differs only from $y_{\mu _{\infty }}(N)$ in the initial and final
consecutive terms, $\sum_{t=1}^{t=t_{0}}\left\vert x_{t}\right\vert $ and $%
\sum_{t=N}^{t=M}\left\vert x_{t}\right\vert $, respectively, where $t_{0}$
is the time (shown in Fig. 2) at which the position $x_{0}$ is visited by
the trajectory initiated at the origin $x=0$. When $t_{0}$, $N$, and $M$ are
all powers of $2$ the differences between the sums become simpler and
expressable in terms of $y_{\mu _{\infty }}(N)$. When $N\rightarrow \infty $
and $M\rightarrow \infty $ the difference between them is only a finite term
$y_{\mu _{\infty }}(t_{0})$. Similar properties hold for the equivalent sums
$z_{\mu _{\infty }}(N)$ and $Z_{\mu _{\infty }}(M)$ that take into account
the signs of positions $x_{t}$.

\begin{figure}[tbp]
\centering \includegraphics[width=12cm,angle=0]{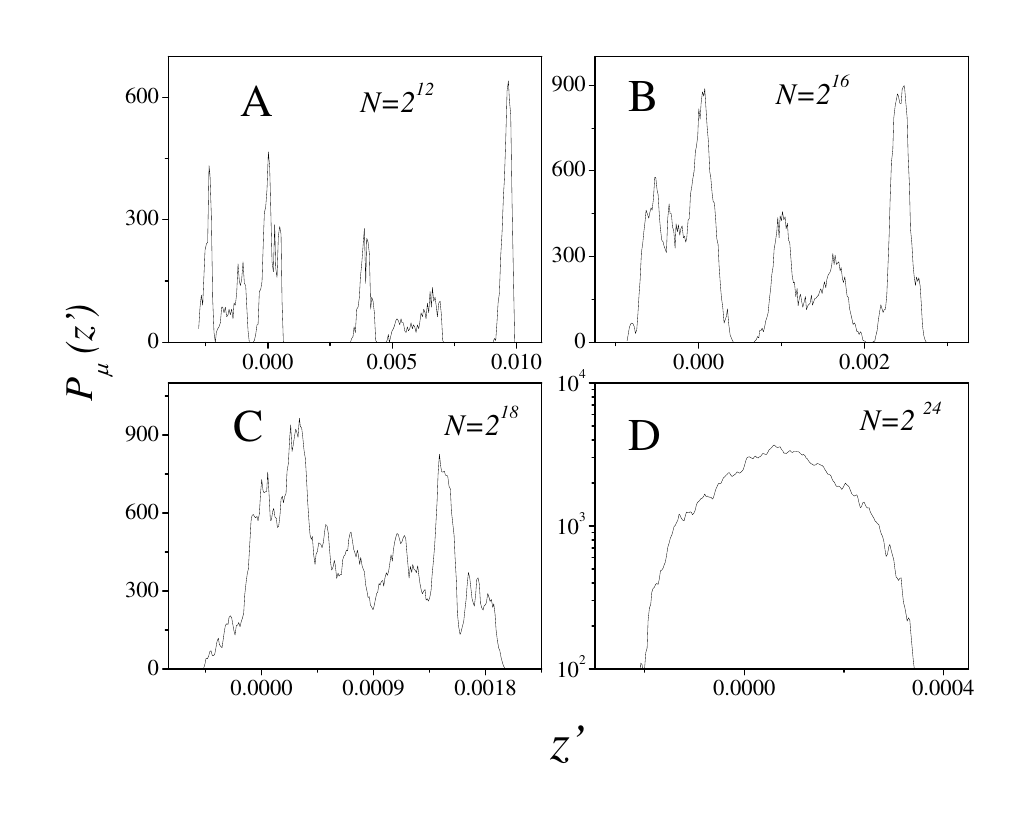}\newline
\caption{Distributions for the sums of positions $x_{t}$, $t=0,...,N$, of a
single trajectory with initial condition $x_{0}=0$ within the $2^{3}$-band
attractor at $\Delta \mu =0.0028448109$. The number of summands $N$ are
indicated in each panel. See text.}
\label{f4}
\end{figure}

\section{Sums of positions for chaotic bands}

We turn now to study the sum of positions of trajectories when $\Delta \mu
\equiv \mu -$ $\mu _{\infty }>0$. We recall that in this case the attractors
are made up of $2^{K}$, $K=1,2,3,...$, bands and that their trajectories
consist of an interband periodic motion of period $2^{K}$ and an intraband
chaotic motion. We evaluated numerically the sums $z_{\mu }(N)$ for a single
trajectory with initial condition $x_{0}=0$ for different values of $\Delta
\mu $. The sum $z_{\mu }^{\prime }(N)$ was then obtained similarly to Eq. (%
\ref{sumabs2}) by substracting the average $\left\langle z_{\mu
}(N)\right\rangle _{x_{0}}$ and rescaling with a factor $N^{-1/2}$. The
panels in Fig. 4 show the evolution of the distributions for increasing
number of summands $N$ for a value of $\Delta \mu $ (chosen for visual
clarity) when the attractor consists of $2^{3}$ chaotic bands. Initially the
distributions are multimodal with disconnected domains, but as $N$ increases
we observe merging of bands and development of a single-domain bell-shaped
distribution that as $N\longrightarrow \infty $ converges in all cases to a
Gaussian distribution. As a check of the ergodic property of chaotic band
attractors we have also evaluated the distributions of sums of positions
starting with an ensemble of uniformly-distributed initial positions $x_{0}$
and obtained results equivalent in all respects to those shown in Fig. 4.
Faster convergence to the Gaussian distribution is achieved in this latter
case.

These numerical results can be understood as follows. We recollect \cite{schuster1, Hu} that the relationship between the number of bands, $2^{K}$, $%
K\gg 1$, of a chaotic attractor and the control parameter distance $\Delta
\mu $ at which it is located is given by $2^{K}\sim \Delta \mu ^{-\kappa }$,
$\kappa =\ln 2/\ln \delta _{F}$, where $\delta _{F}=$ $0.46692...$ is the
universal constant that measures both the rate of convergence of the values
of $\mu $ at period doublings or at band splittings to $\mu _{\infty }$. For
$\Delta \mu $ small and fixed, the sum of sequential positions of the
trajectory initiated at $x_{0}=0$, Eq. (\ref{sumabs1}), exhibits two
different growth regimes as the total time $N$ increases. To specify them we
introduce the difference in value $\delta x_{t}\equiv x_{t}(\mu )-\overline{x%
}_{t}(\mu )$ between the position at time $t$ and the average position
within the band $\overline{x}_{t}(\mu )$ occupied at time $t$. Clearly, when
$K\gg 1$ the average positions $\overline{x}_{t}(\mu )$ approximate the
multifractal positions $x_{t}(\mu _{\infty })$ for $t\leq 2^{K}$. In the
first regime, when $N\ll 2^{K}$, the properties of the sum $z_{\mu }(N)$ do
not differ qualitatively from those of $z_{\mu _{\infty }}(N)$. This is
because the fine structure of the Feigenbaum attractor is not suppressed by
the fluctuations $\delta x_{t}$, as these contribute to the sum individually
during the first cycle of the interband periodic motion. The discrete
multi-scale nature of the distribution for $\mu _{\infty }$ is preserved
when the interband motion governs the sum $z_{\mu }(N)$. The distributions
for $z_{\mu }^{\prime }(N)$ and $z_{\mu _{\infty }}^{\prime }(N)$ are
indistinguishable. In the second regime, when $N\gg 2^{K}$, the situation is
opposite, after many interband cycles the fluctuations $\delta x_{t}$ add up
in the sum and progressively wipe up the fine structure of the Feigenbaum
attractor, leading to merging of bands and to the dominance of the
fluctuating intraband motion. Ultimately, as $N\longrightarrow \infty $ the
evolution of the distribution is similar to the action of the Central Limit
Theorem and leads to a Gaussian stationary end result. It is also evident
that as $\Delta \mu $ increases the first regime is shortened at the expense
of the second, whereas when $\Delta \mu \longrightarrow 0$ the converse is
the case. Therefore there exists an unambiguous $\Delta \mu $-dependent
crossover behavior between the two radically different types of stationary
distributions. This crossover is set out when the $\delta x_{t}$
fluctuations begin removing the band structure in $z_{\mu }^{\prime }(N)$
and ends when these fluctuations have broadened and merged all the chaotic
bands and $z_{\mu }^{\prime }(N)$ forms a single continuous interval. When $%
\mu =\mu _{\infty }$ this process never takes place.

\section{An RG approach for sums of positions}

We are in a position now to put together the numerical and analytical
information presented above into the general framework of the RG approach.
As known, this method was designed to characterize families of systems
containing amongst their many individual states (or in this case attractors)
a few exceptional ones with scale invariant properties and common to all
systems in the family. We recall \cite{fisher1} that in the language of a
minimal RG scheme there are two fixed points, each of which can be reached
by the repeated application of a suitable transformation of the system's
main defining property. One of the fixed points, is termed trivial and is
reached via the RG transformation for almost all initial settings. i.e. for
all systems in the family when at least one of a small set of variables,
named relevant variables, is nonzero. To reach the other fixed point, termed
nontrivial, it is necessary that the relevant variables are all set to zero,
and this implies a severely restricted set of initial settings that ensure
such critical RG paths. The nontrivial fixed point embodies the scale
invariant properties of the exceptional state that occurs in each system in
the family and defines a universality class, while the differences amongst
the individual systems are distinguished through a large set of so-called
irrelevant variables. The variables in the latter set gradually vanish as
the RG transformation is applied to a system that evolves toward the
nontrivial fixed point. Further, when any system in the family is given a
nonzero but sufficiently small value to (one or more of) the relevant
variables, the RG transformation converts behavior similar to that of the
nontrivial fixed point into that resembling the trivial fixed point through
a well-defined crossover phenomenon.

\subsection{RG transformation and fixed points}

The recognition of the RG framework in the properties of the sums of
positions of trajectories in unimodal maps and their associated
distributions is straightforward. It can be concluded right away that in
this problem (as defined here) there is only one relevant variable, the
control parameter difference $\Delta \mu $. There is an infinite number of
irrelevant variables, those that specify the differences between all
possible unimodal maps (with quadratic maximum) and the Feigenbaum map $g(x)$%
. The RG transformation consists of the increment of one or more summands in
the sum (\ref{sumabs1}) followed by centering like in Eq. (\ref{sumabs2}).
The effect of the transformation in the distribution of the sum is then
recorded . For sums of independent variables the transformation is
equivalent to the convolution of distributions \cite{note1}. Notice that the
transformation\ involves no scaling for the sums of positions at $\mu
_{\infty }$. Examination of either Fig. 1 or Fig. 3 indicates that the
values of the sums $y_{\mu _{\infty }}^{\prime }(N)$ and $z_{\mu _{\infty
}}^{\prime }(N)$ are contained within a band of fixed width for all $N$ and
therefore there should not be any scaling for such sums. The Feigenbaum
attractor is not chaotic and its positions $x_{t}$ are not random variables.
On the contrary the values of the sums $y_{\mu }^{\prime }(N)$ and $z_{\mu
_{\infty }}^{\prime }(N)$ with $\mu >\mu _{\infty }$ spread as $N$ increases
and scaling with a factor $N^{-1/2}$ maintains their values contained for $%
N\gg 1$. In this case the $x_{t}$ behave as independent random variables.
There are two fixed-point distributions, the trivial continuum-space
Gaussian distribution and the nontrivial discrete-space multifractal
distribution (as observed in Figs. 1C, 3C and 3D). As explained above, there
is a distinct crossover link between the two fixed-point distributions. Our
results correspond to the dynamics inside the attractors, however, if the
interest lies in considering only the stationary distribution of sums that
do not contain the transient behavior of trajectories in their way to the
attractor \cite{tsallis1}-\cite{tsallis3} our results are expected to give
the correct answers for this case (see the Summary and discussion).

\subsection{ Crossover via band merging}

In Fig. 5 we show a sector of the chaotic bands for the logistic map $f_{\mu
}(x)$ where we indicate the widths of these bands at the control parameter
values $\widehat{\mu }_{K}$ when they split each into two new bands.
Interestingly, if we assume that for a given value of $K$ the widths of
comparable lengths have equal lengths then these widths can be obtained from
the widths of shortest and longest lengths via a simple scale factor
consisting of an inverse power of $\alpha $. See Fig. 5. This introduces
some degeneracy in the widths that propagates across the band splitting
structure. Specifically, the widths scale now with increasing $K$ according
to a binomial combination of the scaling of those that converge to the most
crowded and most sparse regions of the multifractal attractor at $\mu
_{\infty }$. As seen in Fig. 5 the widths form a Pascal triangle across the
band splitting cascade. The total length $L_{K}$ of such chaotic $2^{K}$%
-band attractor can be immediately evaluated to yield%
\begin{equation}
L_{K}=\left( \alpha ^{-1}+\alpha ^{-2}\right) ^{K}.  \label{lengthK}
\end{equation}%
(See Ref. \cite{robledo2} for a similar evaluation relating to the
supercycle diameters occurring across the bifurcation tree for $\mu <\mu
_{\infty }$).

\begin{figure}[tbp]
\centering \includegraphics[width=12cm,angle=0]{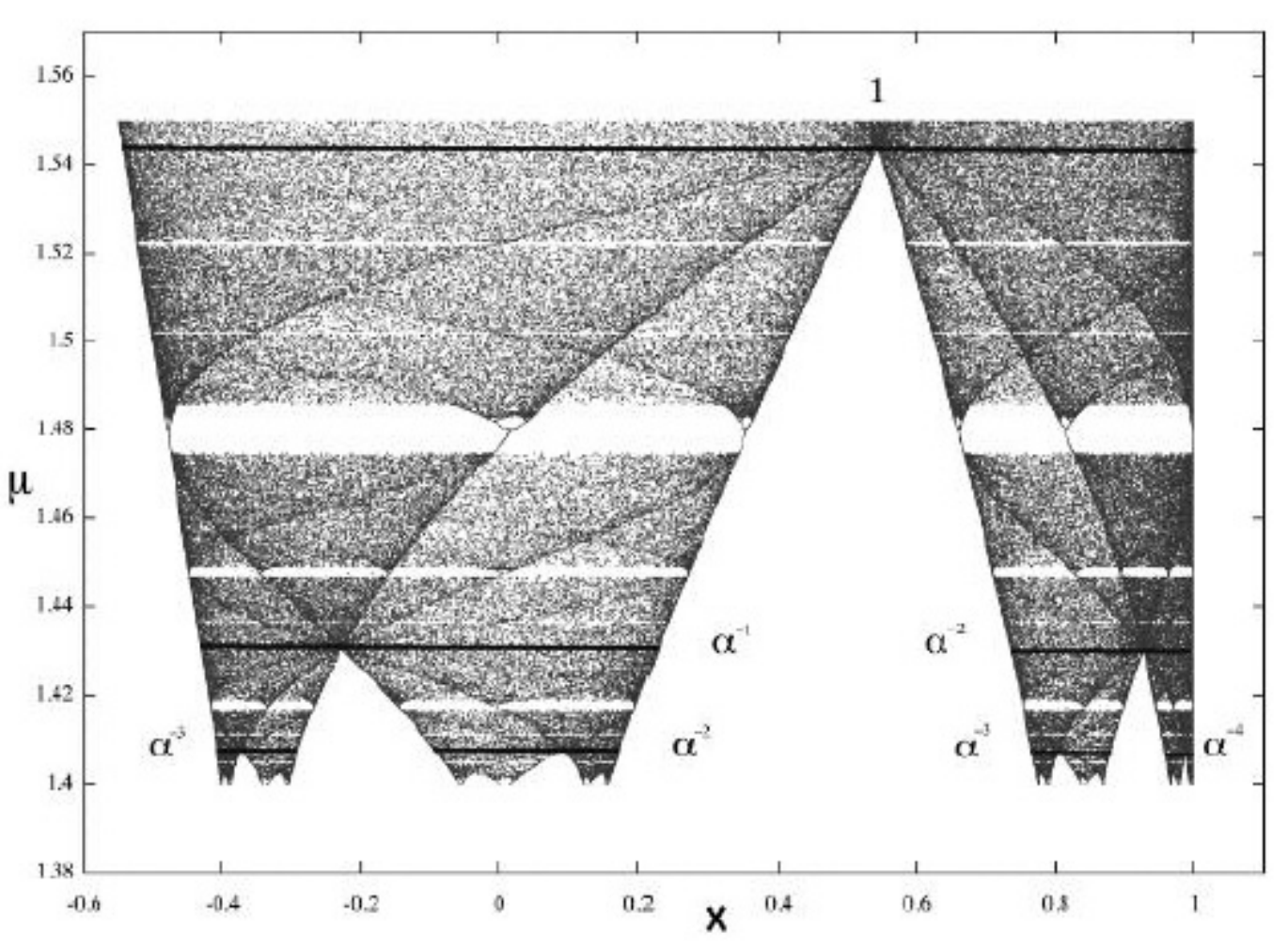}\newline
\caption{Sector of the band splitting cascade for the logistic map $f_{\mu }(x)
$ that shows the formation of a Pascal triangle of band widths (blue lines) at splitting
according to the scaling approximation explained in the text, where $\alpha
\simeq 2.5091$ is the absolute value of Feigenbaum's universal constant.}
\label{f5}
\end{figure}

Now, the sum $z_{\mu }(N)$ can be split into two terms, $z_{\mu }(N)=%
\overline{z}_{\mu }(N)+\delta z_{\mu }(N)$,%
\begin{equation}
\overline{z}_{\mu }(N)=\sum\limits_{t=0}^{N}\overline{x}_{t}\quad,~~~ 
\quad \delta z_{\mu }(N)=\sum\limits_{t=0}^{N}\delta x_{t}. 
\label{sumsz}
\end{equation}%
Similarly to what we have seen for the sums $y_{\mu _{\infty }}(N)$ and $%
z_{\mu _{\infty }}(N)$, the first term $\overline{z}_{\mu }(N)$ is made of a
narrow band of fixed width that shifts altogether linearly with $N$ to
larger values while the band consists of a pattern of period $2^{K}$.
Clearly $\overline{z}_{\mu }(N)$ does not participate in the band merging
process, it is the second term $\delta z_{\mu }(N)$ that fluctuates and
accomplishes band merging. Considering that all the correlated motion of $%
x_{t}$ has been taken over by $\overline{z}_{\mu }(N)$ the fluctuations of $%
\delta z_{\mu }(N)$ correspond to those of independent variables and band
enlargement is measured by the mean square root displacement
\begin{equation}
\left\langle \left[ \delta z_{\mu }(N)\right] ^{2}\right\rangle ^{1/2}\sim
N^{1/2}.  \label{msqrt1}
\end{equation}%
We can estimate the number of summands $N_{K}$ necessary to achieve the
merging of $2^{K}$ bands into a single one by matching the two lengths $%
\left\langle \left[ \delta z_{\mu }(N_{K})\right] ^{2}\right\rangle ^{1/2}$
and $1-L_{K}$. From Eqs. (\ref{lengthK}) and (\ref{msqrt1}) we obtain%
\begin{equation}
N_{K}\sim \left( \alpha ^{-1}+\alpha ^{-2}\right) ^{K}\left[ 1-\left( \alpha
^{-1}+\alpha ^{-2}\right) \right] ,  \label{match1}
\end{equation}%
and considering $N_{K}$ to be of the form $N_{K}=2^{n_{K}}$ with $K\gg 1$ we
obtain
\begin{equation}
n_{K}\sim 2K\ln \left[ \left( \alpha ^{-1}+\alpha ^{-2}\right) /2\right] .
\label{nKabout2K}
\end{equation}%
The crossover estimate in Eq. (\ref{nKabout2K}) for multiband into
single-band distributions coincides in order of magnitude with the number of
summands found necessary in Refs. \cite{tsallis2} and \cite{tsallis3} for
numerical observations of long-tailed distributions resembling the so-called
$q$-Gaussians \cite{tsallis2}.

\section{Summary and discussion}

In summary, we have found that the stationary distribution of the sum of
iterate positions within the Feigenbaum attractor has a multifractal
structure stamped by that of the initial multifractal set, while that
involving sums of positions within the attractors composed of $2^{K}$
chaotic bands is the Gaussian distribution. We considered only the
properties of a single sequence. At this transition the dynamics is
nonergodic and nonmixing, therefore the two natural options, (i) time
average with fixed initial condition and (ii) ensemble average of initial
conditions at a large fixed time, are non- equivalent. A third option is to
perform both ensemble and time averages. We chose option (i) because all
sums with fixed initial positions within the attractor are simply related to
each other via a deterministic term (as explained in Section II). Also,
option (i) allowed us to obtain analytical results in closed form. For the
ergodic chaotic band attractors the choice of a single initial condition
leads to the same result as the use of an ensemble.

In Refs. \cite{tsallis1}-\cite{tsallis3} sums of subsequent values of
trajectories with uniformly distributed initial values across $-1\leq x\leq
1 $ were computed, and their properties were studied after discarding long
transients. In Ref. \cite{tsallis3} transients were\ discarded of lengths
ranging from 2048 to 65536 and their results are reported to be insensitive
to the transient length. A cursory inspection of Ref. \cite{robledo2} (with
regards to Figs. 14 to 18 and the text associated to them) makes it evident
that after iteration times smaller than the smaller transient discarded in
Ref. \cite{tsallis3} the trajectories considered have for all purposes
fallen into the attractor. Thus, these sums are exceptionally well
reproduced by sums of subsequent values $x_{t}$ with initial values inside
the attractor. Next, when it is taken into account that sums belonging to
every such initial value are simply related to that with $x_{0}=0$, it
follows that the choice of sums we considered here capture the limiting
properties of the sums studied in \cite{tsallis1}-\cite{tsallis3}.
The numerical results that approximate $q$-Gaussians and/or L\'{e}vy distributions in \cite{tsallis1}-\cite{grassberger1}  may be understood by observing that the power laws in the $\Delta \mu=0$
multifractal distribution (see Fig. 3) are preserved at the tails of the distributions throughout the band-merging crossover. Only in the limit $N \rightarrow \infty$ the distribution
tails become truly exponential.

We have also shown that the entire problem can be couched in the language of
the RG formalism \cite{note1} in a way that makes clear the identification
of the existing stationary distributions and the manner in which they are
reached. These basic features suggest a degree of universality, and
therefore limited to the critical attractor under consideration, in the
properties of sums of deterministic variables at the transitions to chaos.
Namely, the sums of positions of memory-retaining trajectories evolving
under a vanishing Lyapunov exponent appear to preserve the particular
features of the multifractal critical attractor under examination. Thus we
expect that varying the degree of nonlinearity of a unimodal map would
affect the scaling properties of time averages of trajectory positions at
the period doubling transition to chaos, or alternatively, that the
consideration of a different route to chaos, such as the quasiperiodic
route, would lead to different scaling properties of comparable time
averages. For instance, the known dependence of the universal constant $%
\alpha $ on the degree of nonlinearity $\varsigma $ of a unimodal map would
show as a $\varsigma $-dependent exponents $s$ and $s^{\prime }$ that
control the scale invariant property of the sums of trajectory positions
with $x_{0}=0$ (shown in Figs. 1C, 3C and 3D).

It is worthwhile expanding here our previous comment about the applicability of
our method to other maps. For instance, the stationary distributions associated
to the prototypical circle map \cite{schuster1,Hu} could be determined similarly, thus
extending our study to the route to chaos via quasi-periodicity. The dynamics at
the classic golden mean critical attractor of the circle map exhibits counterparts
with the Feigenbaum attractor concerning all the basic features that we have
made use of here \cite{hernandez}. This can be corroborated via comparison of Fig.
2 with Fig. 3 of \cite{hernandez} and related text therein.

We have contributed to clarify the nature and the circumstances under which
a stationary distribution with universal properties (in the RG sense) may
arise from sums of deterministic variables at the transition between regular
and chaotic behavior, such as those studied here for variables evolving at
zero Lyapunov exponent. In the absence of the fluctuating element present in
chaotic attractors the distribution of the sums of these variables remains
defined on a discrete multifractal set and is kept away from becoming a
known (Gaussian or otherwise) continuum-space limit distribution for random
variables.

\textbf{Acknowledgements.} We appreciate partial financial support by
DGAPA-UNAM and CONACYT (Mexican agencies). AR is grateful for hospitality
received at the SFI.


\begin{thebibliography}{99}
\bibitem{note1} A helpful note about the differences between the
original RG transformation for unimodal maps \cite{schuster1, Hu} and that
presented here may be useful for the casual reader. (i) Feigenbaum's RG
transformation consists of functional composition (and fixed scaling with
the universal constant $\alpha $), while ours is summation of variables
(with no scaling at the transition and scaling with the inverse square root
of total number of terms beyond the transition). (ii) The fixed points in
Feigenbaum's RG are maps, while the fixed points in ours are distributions.
(iii) Feigenbaum's RG generates the period doubling cascade, while ours
produces band merging. (iv) The purpose of Feigenbaum's RG is to
characterize the critical attractor, while ours is to characterize
stationary distributions.
\end{thebibliography}
\end{document}